\documentclass[preprint,12pt]{article}

\usepackage{amssymb}
\usepackage{adjustbox}
\usepackage{pgf,pgfarrows,pgfnodes,pgfautomata,pgfheaps}
\usepackage{tikz}
\usepackage[all]{xy}
\usetikzlibrary{matrix}
\usepackage{graphicx}
\usepackage{multimedia}
\usepackage{graphics}
\usepackage{amssymb,amsmath,amsfonts,graphicx,tabularx}
\usepackage{tikz-cd}
\usepackage{soul}
\usepackage{xcolor}
\usepackage{ulem}
\usepackage{comment}
\usepackage{float}

\begin{document}

\title{Parameterizing density operators with arbitrary symmetries to gain advantage in quantum state estimation}

\author{{\sc In\'{e}s Corte} $^{1,2}$ \ {\sc ,} \ {\sc Marcelo Losada}$^{3}$ \ {\sc ,} \ {\sc Diego Tielas}$^{1,2}$ \ {\sc ,}\\ \ {\sc Federico Holik}$^{1}$ \ {\sc and} \ {\sc Lorena Reb\'{o}n}$^{1,2}$}

\maketitle

\begin{center}

\begin{small}
1- Instituto de Física La Plata (CONICET-UNLP), Calle 113 entre 64 y 64 S/N, 1900, La Plata, Buenos Aires, Argentina \\

2- Departamento de Ciencias Básicas, Facultad de Ingenier\'ia, Universidad Nacional de La Plata, Calle 48 y 116, 1900, La Plata, Buenos Aires, Argentina \\

3- Facultad de Matemática, Astronomía, Física y Computación, Universidad Nacional de Córdoba, Av. Medina Allende S/N, Ciudad Universitaria, Córdoba, X5000, Córdoba, Argentina

\end{small}
\end{center}

\vspace{1cm}

\begin{abstract}
In this work, we show how to parameterize a density matrix that has an arbitrary symmetry, knowing the generators of the Lie algebra (if the symmetry group is a connected Lie group) or the generators of its underlying group (in case it is finite). This allows to pose MaxEnt and MaxLik estimation techniques as convex optimization problems with a substantial reduction in the number of parameters of the function involved. This implies that, apart from a computational advantage due to the fact that the optimization is performed in a reduced space, the amount of experimental data needed for a good estimation of the density matrix can be reduced as well. In addition, we run numerical experiments and apply these parameterizations to quantum state estimation of states with different symmetries.
\end{abstract}

\section{Introduction}

Quantum state estimation lies at the heart of many quantum information tasks, variational quantum algorithms and certification of quantum devices. However, the problem of  
estimating unknown quantum states is generally hard because the number of independent parameters needed for a full description grows exponentially with the number of qubits. Thus, many techniques have been developed for particular classes of states that satisfy certain conditions, which imposes constraints that reduce the number of independent parameters needed for a full specification of the state. An important example of such conditions is the existence of physical symmetries in the generated states. In some cases, symmetries have their origin in the physical properties of the system under consideration, as is the case of the permutational invariance of identical particle systems \cite{Boson_Sampling,Joshi_2021}. In other cases, symmetric states can be generated intentionally to carry out certain specific tasks \cite{Pollatesek-2004,Ouyang-2014}.

Recently, a method for estimating quantum states with arbitrary symmetries using the MaxEnt technique was developed in \cite{MaxEntSolutions,tielas2021performance}. One of the main advantages of this method is that one obtains a substantial reduction in the number of observables needed to perform a reasonable estimation of the state. However, it has the drawback that the numerical optimization is carried out over the whole space of density operators. Thus, while the number of measurements is substantially reduced, the fact that symmetric states need less parameters is not used to take advantage during the numerical computation.

Another example of a relevant method for quantum state estimation that relies on symmetries is the permutationally invariant quantum tomography \cite{PermutationallyInvariantQT}. This method extracts the permutationally invariant part of an unknown quantum state, which has much less independent parameters than the full state. In fact, the number of measurements needed to determine the permutationally invariant part of an unknown state grows polynomially with the number of qubits \cite{Klimov-Permutational-POVM}. The formulation of the problem and the choice of the observables to measure in laboratory implementations depends on the specific properties of the symmetry group \cite{PermutationallyInvariantQT,Moroder_2012}. Up to now, there is no similar method for other symmetries.

In conclusion, a reduction in the computational complexity of the problem can be attained expressing a density operator that has the given symmetry as a combination of elements of a certain symmetric subspace of the space of Hermitian matrices. Generalizing the method of permutationally invariant tomography, the aim of this work is to reformulate the task of state estimation with arbitrary symmetries as a non-linear convex optimization problem. For that, we present a method for writing a quantum state with a given symmetry as a linear combination of the elements of a proper subspace of the space of Hermitian matrices. In doing so, we obtain a reduction in the dimensionality of the resulting convex optimization problem. This strategy has the following advantages with regards to previous works:
\begin{itemize}
\item It works for a very general family of symmetries, namely those associated with finite groups and connected Lie groups. Almost all symmetries of interest in physics fall into these two categories. As relevant examples, we can mention: arbitrary rotations, permutationally invariant states (and all its Bosonic and Fermionic subfamilies) and Werner states.
\item Our method only depends on the mathematical peculiarities of the symmetry group via its generators. No further dependence on the particular properties of the group are needed. This means that, in practice, the main problem consists in identifying a set of generators of the Lie algebra (for connected Lie groups) or a set of generators of the group (for finite groups). In most tasks, this is equivalent to the mere specification of the group involved\footnote{For example, when we have a certain rotation, say a rotation in the $\hat{z}$ direction, we express it as $e^{-\frac{i}{\hbar}\sigma_{z}\theta}$, with $\theta\in\mathbb{R}$. Then, the Lie algebra generators are given by the set $\{\sigma_{z}\}$. Werner and permutationally-invariant state generators can be computed easily, as shown in \cite{tielas2021performance}.}. This feature simplifies the mathematical formulation of the problem considerably in comparison with other methods.
\item The empirically-implemented observables need not possess the symmetry of the state. Only a mild ``non-orthogonality" condition is required. Thus, our estimation technique can, in principle, be easily adapted to the particularities of very specific measurement set-ups (a feature that would allow to save a lot of experimental resources in many cases).
\item These features result in a reduction of both the number of independent parameters, that define the dimension of the convex optimization problem, and the number of measured observables.
\item Finally, the above methodology can be formulated as a simple algorithm, so that it can be used in practice without the need of delving deep into the technical details.
\end{itemize}

This paper is organized as follows. In Section \ref{s:Parameterizing} we show how to write a density operator having an arbitrary symmetry as a linear combination of a proper subspace of the space of Hermitian operators. Next, in Section \ref{s:ConvexOpt}, and after analysing the linear inversion method, we show how to get advantage of this decomposition to formulate a convex optimization problem with a reduced number of parameters. We sketch how to define the symmetric part of a state (an approach similar to \cite{Moroder_2012} but considering arbitrary symmetries instead) in Section \ref{s:SymmetricPart}. Then, we test the performance of our method in a more realistic scenario using states with different symmetries in Section \ref{s:Examples}. Finally, in Section \ref{s:Conclusions}, we summarize our results.

\section{Parameterization of a density operator which is symmetric under the action of a group}\label{s:Parameterizing}

In this section we assume that a given density operator $\rho$ possesses a certain symmetry. This symmetry can be described by a connected Lie group, as is the case of continuous rotations and Werner states \cite{Eggeling2001}, or by a finite group, such as discrete rotations with finite elements and permutations. Groups will be denoted by $G$ and their Lie algebras by $\mathfrak{G}$. The set of states that have a certain symmetry form a proper convex subset\footnote{Except when the symmetry group is the trivial one, i.e., $G=\{I\}$} of the convex set of quantum states \cite{Holik-Massri-Plastino-IJGMMP-2016}. In general, the dimension of this proper convex subset will be much lower than that of the original space, as we will see in Section \ref{s:Examples}. In order to use this reduction in the dimensionality, we need to rewrite the original state as a linear combination in a proper subspace of the space of Hermitian operators.

Let us start first showing how to do this for connected Lie groups. We will reproduce some of the results presented in \cite{MaxEntSolutions} to make the presentation more self-contained. In what follows, we consider the scalar product $\langle A;B\rangle = \mbox{tr}(AB^{\dagger})$
in the vector space of complex matrices $\mathbb{C}^{n\times n}$, which is also a Hilbert space. Notice that, if $A$ and $B$ are Hermitian, then $\mbox{tr}(AB^{\dagger})=\mbox{tr}(AB)$. It also defines a real scalar product in the real-Hilbert space of Hermitian matrices.

Given a connected Lie group $G$ with Lie algebra $\mathfrak{G}$, assume that its generators are given by $\{Q_{k}\}_{k=1,\ldots,m}$ (i.e., $\mathfrak{G}$ is the linear span of $\{Q_{k}\}_{k=1,\ldots,m}$). Next, consider a basis for the space of Hermitian matrices: $\{O_{j}\}_{j=1,\ldots,n\times n}$. Then, one can prove that a state $\rho$ that possesses the symmetry ($U\rho U^{\dagger}$ for all $U \in G$) satisfies \cite{MaxEntSolutions}:

\begin{eqnarray}\label{e:ConditionsSymmetry}
	\langle [i Q_k,O_j] \rangle&=&
	\mbox{Tr}\left(i  [Q_k,O_j]\rho\right)=0,\nonumber\\ ~~\forall k =1,\ldots,m,
	&~& \forall \,  j= 1, \ldots, n\times n,
\end{eqnarray}

The reason is the following. Assume that a density operator $\rho$ is invariant under the action of $G$: $U\rho U^{\dagger}=\rho$, for all $U\in G$. Now, given that $U$ is unitary and $G$ is connected, then each $U\in G$ can be written as $U=e^{i M t}$, where $M\in \mathfrak{G}$ ($A$ is Hermitian and belongs to the Lie algebra of $G$) and $t\in\mathbb{R}$ (notice that $t=0$ corresponds to $U=Id$, the unit element of the group). Thus,

\begin{equation}
    e^{iMt}\rho e^{-iMt}=\rho
\end{equation}

\noindent for all $t\in \mathbb{R}$ if and only if

\begin{equation}
     e^{iMt}\rho e^{-iMt}-\rho=0
\end{equation}

\noindent which -- using the expansion of the exponential function -- can rewritten as

\begin{equation}
    (1-iMt)\rho(1+iMt)-\rho+O(t^{2})=-iMt\rho +\rho iMt+O(t^{2})=-it[\rho,M]+O(t^{2})=0
\end{equation}

\noindent The above equation holds for all $t\in\mathbb{R}$ if and only if
\begin{equation}
[\rho,M]=0
\end{equation}

\noindent from which it follows that $U\rho U^{\dagger}=\rho$ for all $U\in G$, if and only if $[\rho,M]=0$ for all $M\in \mathfrak{G}$. But the above condition is true if and only if $[\rho,Q_{k}]=0$ for all the generators of the Lie algebra $\{Q_{k}\}_{k=1,\ldots,m}$.

Equations \eqref{e:ConditionsSymmetry}  can be seen as the kernel of a linear functional $\mbox{Tr}\left(i  [Q_k,O_j]\ldots\right)$, which is a linear subspace and is thus convex. The density operators that satisfy these equations are just contained in the intersection of all the kernels with the convex set of quantum states. Since the intersection of convex sets is a convex set, Eqs. \eqref{e:ConditionsSymmetry} define a convex set $\mathcal{C}_{G}$ of density matrices. This set is also compact \cite{Holik-Massri-Plastino-IJGMMP-2016}. Notice that, under these conditions, any continuous strictly convex (concave) function 
defined over $\mathcal{C}_{G}$ will attain a unique maximum (minimum) element. If we add an arbitrary number of mean value equations of the form $\mbox{tr}(\rho O)=o$ (being $O$ an Hermitian matrix and $o\in \mathbb{R}$), we will also obtain a convex set (with the sole condition that the equations be compatible).

Applying the scalar product, an equation of the form

\begin{equation}
    \mbox{Tr}\left(i  [Q_k,O_j]\rho\right)=0
\end{equation}

\noindent can be rewritten as

\begin{equation}\label{e:Projection}
    \langle\rho;i[Q_k,O_j] \rangle=0.
\end{equation}

\noindent The geometric interpretation of Eq. \eqref{e:Projection} is that the projection of $\rho$ with respect to the $\mathbb{R}$-subspace $\mathbb{T}$, generated by the set $\{i[Q_k,O_j]\}_{k=1,\ldots,m;j=1,\ldots,n\times n}$, is null. Define then $\mathbb{S}_{G}:=\mathbb{T}^{\bot}$ (i.e., $\mathbb{S}_{G}$ is the orthogonal complement of $\mathbb{T}$). We call $\mathbb{S}_{G}$ the \textit{symmetric subspace} associated to the group $G$. Thus, it follows that the projection of $\rho$ is only non-null in the subspace $\mathbb{S}_{G}$. Let $\{S_{1},\ldots,S_{r}\}$ be an orthogonal basis of $\mathbb{S}_{G}$ (we are assuming that $dim(\mathbb{S}_{G})=r$). Thus, by the above considerations, we must have that:

\begin{equation}\label{e:Parameterization}
    \rho = \sum^{r}_{i=1} \alpha_{i}S_{i}
\end{equation}
\noindent for some real parameters $\{\alpha_{i}\}_{i=1,...,r}$, with $r<(2^{N}\times 2^{N})-1$. Exploiting that $\rho$ has a symmetry defined by the action of $G$, we have managed to write it using less parameters than those needed for a not necessarily invariant state. All we needed to generate $\mathbb{S}_{G}$ were the Lie algebra generators $\{Q_{k}\}_{k=1,\ldots,m}$ (or the generators $\{G_{1},\ldots,G_{m}\}$ when the group is finite). No further reliance on other group features is required. Indeed, with these inputs, it is possible to find an orthogonal basis of $\mathbb{S_{G}}$ using only linear algebra operations.

Equation \eqref{e:Parameterization} must be interpreted as follows: all states $\rho$ which are invariant under the action of the group $G$ can be written in that way for some unknown parameters $\alpha_{i}$. Equation \eqref{e:Parameterization} can be used as the starting point of a convex optimization problem: find the optimal values of the $\alpha_{i}$ in such a way that $\rho$ satisfies some constrains (typically, positivity, unit trace, and a set of mean values experimentally obtained) and maximizes (or minimizes) a convex (or concave) function, such as entropy. Notice that the manifold defined by Eq. \eqref{e:Parameterization} is convex, which gives a non-linear convex optimization problem.

The advantage of using such a parameterization is that the set of parameters, for many symmetries, is much smaller than the dimension $d$ of the original space. For example, a permutationally symmetric state $\rho$ is defined by ${d\times d+N-1}\choose{N}$ $-1\sim N^{d^2-1}$ independent real parameters, which exhibit a cubic growth with the number $N$ of qubits. In Section \ref{s:Examples} we give concrete examples of this reduction in the number of parameters for different symmetries.

The construction for finite groups is completely analogous to the one exposed above. Let $G$ be a finite group generated by a set $G_{0}=\{g_{1},g_{1},\ldots,g_{n}\}$ (for finite groups, the existence of this set follows trivially from the definition of \textit{generators}; see \cite{Lang_Algebra}, page 9). This means that any element $g\in G$ can be written as a (finite) product of elements of $G_{0}$ (and their inverses). In the Hilbert space, the elements of $G_{0}$ are represented by a set of unitary operators $RG_{0}=\{U_{1},U_{2},\ldots,U_{n}\}$, that, again, generate a representation $RG$ of $G$ in a similar way because any representation establishes a group homeomorphism. Now, the invariance condition of $\rho$ can be stated as

\begin{equation}
U \rho U^{\dagger} = \rho.
\label{e:19}
\end{equation}


\noindent It is straightforward to prove that Eq. \eqref{e:19} is equivalent to

\begin{equation}\label{e:FiniteSymmetry}
U \rho U^{\dagger} = \rho\,\,,\,\,\forall U \in RG_{0}.
\end{equation}

\noindent Using Eq. \eqref{e:FiniteSymmetry} and proceeding similarly to the non-discrete case, we obtain that $\rho$ possesses the desired symmetry if and only if $[\rho,U_{k}]=0$ for all $k=1,...,n$. From these equations, we can proceed in a way completely analogous to the continuous case.

\section{State estimation using decompositions with fewer parameters}\label{s:ConvexOpt}

In this section we discuss how to use the decomposition in Eq. \eqref{e:Parameterization} to determine a symmetric -- but otherwise unknown -- quantum state. Even though our results will mainly focus on the convex optimization application, we first show how to such decomposition for the case of linear inversion.

\subsection{Linear inversion}

Several reconstruction techniques are based on direct linear inversion of experimental data. This method is conceptually simple and straightforward. But it might yield an unphysical representation of the state, since positivity is not guaranteed. However, under certain conditions, linear inversion is still relevant in many circumstances of practical interest \cite{Kaznady-LinearInversion}.

Assuming that the unknown state has a symmetry with an underlying group $G$, we use the decomposition in Eq. \eqref{e:Parameterization} in the Born rule to obtain the expectation value equation for an observable $A$:


\begin{equation}\label{e:SymmetricMeanValue}
    \langle A\rangle = \mbox{tr}(\rho A)= \sum^{r}_{i=1} \alpha_{i}\mbox{tr}(S_{i} A)
\end{equation}

\noindent The left hand side of Eq. \eqref{e:SymmetricMeanValue} represents the experimentally acquired mean value of $A$\footnote{In the ideal situation in which experimental noise is reduced to zero, the theoretical and empirical numbers would be equal.}. The right hand side represents our theoretical representation of the state under the symmetry assumptions. Notice that, if $\mbox{tr}(S_{i} A)=0$ for all $i$, then, we would obtain an equation of the form $0 = 0$. Thus, in the case in which $A$ is orthogonal to $\mathbb{S}_{G}$, the content of Eq. \eqref{e:SymmetricMeanValue} is vacuous. On the contrary, if for some $i$ we have $\mbox{tr}(S_{i} A)\neq 0$, it might give place to a linearly independent equation, and then, it will give us non-trivial information about some of the $\alpha_{i}$'s. Thus, assume that we measure a family of observables $\{A_{i}\}_{i=1,...,r}$ satisfying that  $\mbox{tr}(S_{j}A_{i})\neq 0$ for all $i$ and at least for some $j$. In that case, we obtain $r$ equations:

\begin{align}
    \langle A_{1}\rangle &= \sum^{r}_{i=1} \alpha_{i}\mbox{tr}(S_{i} A_{1})\nonumber\\
 \langle A_{2}\rangle &= \sum^{r}_{i=1} \alpha_{i}\mbox{tr}(S_{i} A_{2})\nonumber\\
 &\vdots&\nonumber\\
 \langle A_{r}\rangle &= \sum^{r}_{i=1} \alpha_{i}\mbox{tr}(S_{i} A_{r})\nonumber
\end{align}

\noindent from which we can obtain the $r$ unknown coefficients, and therefore a representation of the state. Notice that if we did not use this information about the symmetry, the number of unknown parameters would be of the order of $2^{N}\times 2^{N}$ (with a concomitant increment in the number of experimental observables needed). This illustrates the reduction in the number of experimental observables and computational complexity obtained using decomposition (\ref{e:Parameterization}).

In some specific situations, and with suitable modifications, the linear inversion approach can be an effective tool for all practical purposes \cite{Kaznady-LinearInversion}. Another relevant feature of the above procedure is that the sole condition on the $A_{i}$'s is that they are not orthogonal to $\mathbb{S}_{G}$ (which is a quite mild restriction). Due to the freedom in choosing the $A_{i}$'s, this method can be adapted for each experimental setup, considering observables that can be easily implemented in it. 

\subsection{Quantum state estimation as an optimization problem}

Now, we turn to the general problem of tomographic schemes based on optimization techniques under the premise that the unknown density operator possesses a symmetry defined by $G$. This methodology lies at the basis of many quantum state estimation protocols, such as MaxLix \cite{Moroder_2012}, MaxEnt \cite{tielas2021performance}, MaxLik-MaxEnt \cite{Teo-2011}, and Variational convex optimization \cite{Goncalves-2013-MaxEntTomography}.

Let $F$ be a continuous convex function. Given $G$, we define $\mathbb{S}_{G}$ as in Section \ref{s:Parameterizing}. Let $F(\rho)=S(\alpha_{1},\ldots,\alpha_{r})$ be a function with a defined concavity/convexity. Maximize (or minimize, depending on the case) $F(\alpha_{1},\ldots,\alpha_{r})$
under the condition that

\begin{equation}
    \rho=\sum^{r}_{i=1} \alpha_{i}S_{i}\succeq 0,
\label{e:rho-parametrization}
\end{equation}

\noindent the normalization condition $\mbox{tr}(\rho)=1$, and eventually, a set of extra constrains, such as

\begin{align}
     \mbox{tr}(\rho A_{1})=a_{1} \\ \nonumber
    \vdots \\ \nonumber
        \mbox{tr}(\rho A_{r})=a_{r}
\end{align}

\noindent where $A_{1}$, $A_{2}$, $\ldots$, $A_{r}$ are Hermitian operators and $a_{1}$, $a_{2}$, $\ldots$, $a_{c}$ are real are their respective mean values. For the MaxLik problem, the mean values of a suitably chosen set of projection operators -- or more generally, POVMs -- are introduced as parameters of the functional. As we discussed in the previous section, it is requested that they form a quorum set or otherwise the solution will not be unique. The MaxEnt, MaxEnt-MaxLik and Variational techniques allow for an even stronger reduction in the number of measured observables, while keeping a reasonable quality in the estimation output \cite{tielas2021performance}.

\section{Symmetric part of a density operator (for an arbitrary symmetry)}\label{s:SymmetricPart}

From the above considerations, it follows that, given a group $G$, we can split the vector space of $n\times n$ complex matrices in two orthogonal subspaces, namely, $\mathbb{S}_{G}$ and $\mathbb{S}_{G}^{\bot}$. Thus, it should be obvious that an arbitrary density operator $\rho$ can be, in principle, written as:

\begin{equation}
    \rho = \rho_{G^{\bot}} + \rho_{G},
\end{equation}

\noindent where

\begin{equation}
    \rho_{G} = \sum^{\mbox{dim}(\mathbb{S})}_{i=1}\mbox{tr}(\rho S_{i}) S_{i}
\end{equation}

\noindent and

\begin{equation}
    \rho_{G^{\bot}} = \sum^{\mbox{dim}(\mathbb{S}^{\bot})}_{i=1}\mbox{tr}(\rho S^{\bot}_{i}) S^{\bot}_{i}
\end{equation}

\noindent where the $S^{\bot}_{i}$'s form an orthonormal basis of $\mathbb{S}^{\bot}$. Its particular form is of no use for our estimation scheme.

Under some circumstances, finding the symmetric part of an unknown density operator can be useful for quantum information theory (see for example the arguments presented in \cite{PermutationallyInvariantQT} and \cite{Moroder_2012} for the case of permutationally invariant states). Here, we have shown how to generalize the symmetric part of a density operator to a much larger family of symmetries.
If we now chose a collection of linearly independent observables $\{A_{i}\}_{i=1,\:\ldots,\:\mbox{dim}(\mathbb{S})}$ that can be written as linear combinations of elements of $\mathbb{S}$, we can perform a tomographic scheme that estimates the symmetric part of an unknown density operator for an arbitrary symmetry defined by $G$.

\section{Examples and numerical simulations}\label{s:Examples}
In this section, we apply the ideas exposed above to estimate multi-qubit quantum states with specific symmetries. To this end, we combine the parameterization from Section \ref{s:ConvexOpt} with the variational quantum tomography (VQT) technique described in Refs. \cite{Goncalves-2013-MaxEntTomography, maciel2011variational}. We refer to this hybrid approach as \textit{group-invariant tomography} (GIT). We will study scenarios with a number of measurements equal to the number of parameters of the symmetric state involved. Nevertheless, as our algorithm is based on VQT, it can be used even when the number of parameters is less than quorum \cite{Goncalves-2013-MaxEntTomography}, as in the case of MaxEnt estimation.
Our analysis is focused on permutationally invariant states, Werner states, and states which are invariant under global rotations or under individual qubit rotations.

\begin{figure}[h!]
\centering
\includegraphics[width=0.75\textwidth]{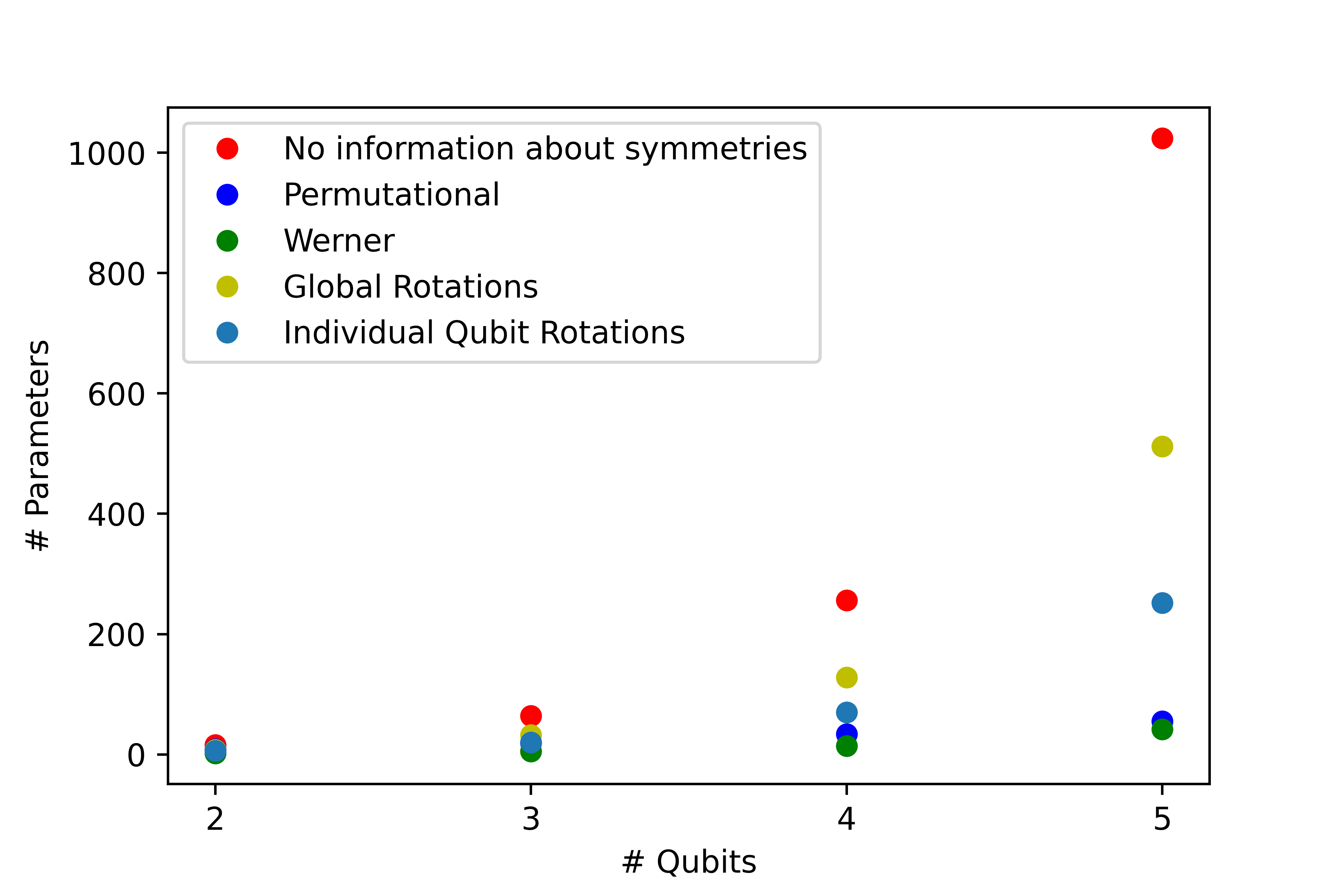}
\caption{Number of independent parameters in the decomposition of Eq.~\eqref{e:rho-parametrization} as a function of the number of qubits for different symmetries. If no symmetries of the state are considered (red dots), the number of parameters grows exponentially. Fewer parameters usually imply fewer measurements in experimental setups, and a computational problem of smaller dimension.}
\label{f:Dimensions}
\end{figure}

In Fig. \ref{f:Dimensions}, we show the number of independent parameters in the symmetric part of a density operator with the given symmetry as a function of the number of qubits.
For comparison, we also include the number of parameters required to estimate a multi-qubit state via \textit{standard quantum tomography} (SQT).

For an approach to the performance of the GIT in a real experimental scenario, in the following sections we
use such a scheme to reconstruct quantum states with specific symmetries by considering simulated measurement results that could be obtained in hypothetical experiments. The simulated lab frequencies were generated modifying the theoretically predicted probabilities with a binomial or Gaussian probability distribution. As figure of merit, we compute the fidelity of reconstruction between the resulting estimated state and the ideal target state.

The GIT scheme was implemented algorithmically in Python following the ideas from Section \ref{s:Parameterizing}. First, it computes the commutators $[Q_i, O_j]$ for the given symmetry, where $Q_i$ is a Lie Generator and $O_j$ is a basis element of the space of hermitian $2^N \times 2^N$ matrices. Then, it uses Gram-Schmidt orthogonalization (GSO) to determine a subset of $c$ linearly independent commutators. This set of $c$ elements is completed with $\{O_j\}_{j=1,\dots,2^{2N}}$ and, from this new set of $c+2^{2N}$ ordered elements, a subset of $2^{2N}$ orthogonal elements is obtained applying again GSO. The first $c$ elements generate $\mathbb{S}_G^{\bot}$, implying that all remaining elements form an orthogonal basis of $\mathbb{S}_G$. Thus, these $r=2^{2N} - c$ matrices are the $S_i$'s from Eq.~\eqref{e:Parameterization}. Finally, this expression of $\rho$ is integrated into the VQT formulation \cite{Goncalves-2013-MaxEntTomography, maciel2011variational}, that we reproduce here:

\begin{eqnarray}\label{e:OptimizationGoncalves}
    \min_{\rho,\,\Delta} \sum_{i\in\mathcal{I}} \Delta_{i} + \sum_{i\notin \mathcal{I}}\mbox{tr}(E_{i}\rho)\\\nonumber
    \mbox{subject to}\,\,| \mbox{tr}(E_{i}\rho)-f_{i}|\leq \Delta_{i}f_{i}, \,\,\, i \in\mathcal{I}\\\nonumber
    \Delta_{i}\geq 0\\\nonumber
    \mbox{tr}(\rho)=1\\\nonumber
    \rho \succeq 0
\end{eqnarray}

\noindent where $\{E_i\}$ is the measurement set, $\{f_i\}$ the measured frequencies (or mean values), $\{\Delta_i\}$ the
tolerances, and $\mathcal{I}$ stands for the set of indexes of measured
data. We will restrict to scenarios in which the number of measurements equals the number of parameters in the parameterization (but the VQT technique still works reasonably with fewer measurements). In most examples below, the $E_{i}$'s will be projectors and the corresponding $f_{i}$'s will be frequencies (i.e., numbers in the interval $[0,1]$). In the Werner case, we also considered the $S_{i}$'s, which are Hermitian but not necessarily projective. Since the $S_{i}$'s are not necessarily POVM's, some $f_{i}$'s might be negative. Accordingly, we replaced the second restriction of Eqn. \ref{e:OptimizationGoncalves} with $| \mbox{tr}(E_{i}\rho)-f_{i}|\leq \Delta_{i}|f_{i}|$ in the simulations. In our code, the convex optimization problem defined by Eqn. \ref{e:OptimizationGoncalves} is solved using the CVXOPT package \cite{CVXOPT}.

\subsection{Permutationally Invariant States}\label{ejemplo:PI}

We will start
by considering a quantum source that generates permutationally invariant states, which are associated with a discrete symmetry. For example, in the case of pure states,
these are symmetric or antisymmetric under the exchange of any two particles of the system.

Let us briefly describe the generators of the symmetry group. For an $N$-qubit state, $|\psi\rangle=|\psi_{1}\rangle\otimes\ldots\otimes|\psi_{i}\rangle\otimes\ldots\otimes|\psi_{j}\rangle\otimes\ldots\otimes|\psi_{N}\rangle$, the action of the permutation operator $P_{ij}$ ( $\forall i,j=1,\ldots,N$, $i\neq j$) is given by
\begin{equation}\label{permutator_operattor}
P_{ij}|\psi\rangle=|\psi_{1}\rangle\otimes\ldots\otimes|\psi_{j}\rangle\otimes\ldots\otimes|\psi_{i}\rangle\otimes\ldots\otimes|\psi_{N}\rangle,
\end{equation}
Thus, for arbitrary $(i,j)$, a permutationally invariant state $|\psi\rangle$ must satisfy the relation $P_{ij}|\psi\rangle=\pm|\psi\rangle$, where ``$+$" stands for Boson-like qubits and ``$-$" for Fermion-like qubits. Furthermore, if the quantum system is described by a density operator $\rho_{\mathrm{(PI)}}$, the system is said to have permutational symmetry if the following relation is satisfied:
	\begin{eqnarray}
	P_{ij}\rho_{\mathrm{(PI)}} P_{ij}=\rho_{\mathrm{(PI)}}, ~~~\forall~ i,j=1,\dots, N.
	\end{eqnarray}
For an $N$-qubit system, it suffices to consider the generators $\{P_{12},P_{13},P_{14},\ldots,P_{1N}\}$, which can be computed as matrices, for example, in the computational basis. This set of operators is the only input our algorithm needs to compute the orthogonal basis of $\mathbb{S}_G$ appearing in the parameterization of Eq. \eqref{e:Parameterization}.

We ran numerical simulations of the tomographic state task based on a potential quantum optics scenario. In a real experiment, the expectation value of each observable $E_i$ in the target state, $\mathrm{tr}(E_i\rho)$, is approximated by the relative frequency $f_i= n_i/N_{trials}$, where $n_i$ is the number of trials (or pulses) that resulted in a click of the detectors and $N_\text{trials}$ is the total number of trials. These frequencies are related to single-qubit projective measurements on the usual basis of polarization \cite{Kwiat}
$$\left\{ |H\rangle =\left(\begin{array}{c}
    1\\
    0
\end{array} \right), \: |V\rangle =\left(\begin{array}{c}
    0\\
    1
\end{array} \right)\right\},\: \left\{|D\rangle =\frac{1}{\sqrt2}(|H\rangle +|V\rangle),\: |R\rangle =\frac{1}{\sqrt2}(|H\rangle - i|V\rangle)\right\} $$

To simulate these experiments, we have considered the detection noise as the only source of noise and simulated detection frequencies using a binomial distribution $B(N_\text{trials},p_\text{det})$, where $p_\text{det}$ the probability of detecting a photon in a given mode. Since we are dealing with a photonic system, this probability can be modeled as
$$p_\text{det} = 1 - e^{-\mu p_\text{ideal} - \lambda_\text{dc}}, $$
where $\mu$ is the mean photon-number per laser pulse, $p_\text{ideal}$ is the ideal detection probability and $\lambda_\text{dc}$ is the number of dark counts (dc) \cite{Goyeneche2015,rebon19}. In our simulations, we used $\mu = 0.18$, $N_\text{trials} = 5\times10^5$ and different values of $\lambda_\text{dc}$.

\begin{figure}[ht]
\centering
\includegraphics[width=0.75\textwidth]{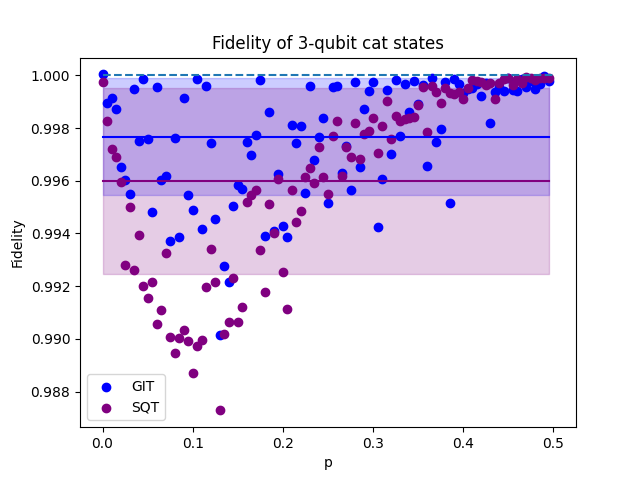}
\caption{Fidelity of 3-qubit cat states estimated via GIT (blue dots) and SQT (purple dots) algorithm from \cite{Kwiat-python}. Horizontal lines and shaded areas represent the mean fidelity value and standard deviation respectively for each tomographic scheme.}
\label{f:3q_Cat}
\end{figure}

\begin{figure}[H]
\centering
\includegraphics[width=0.75\textwidth]{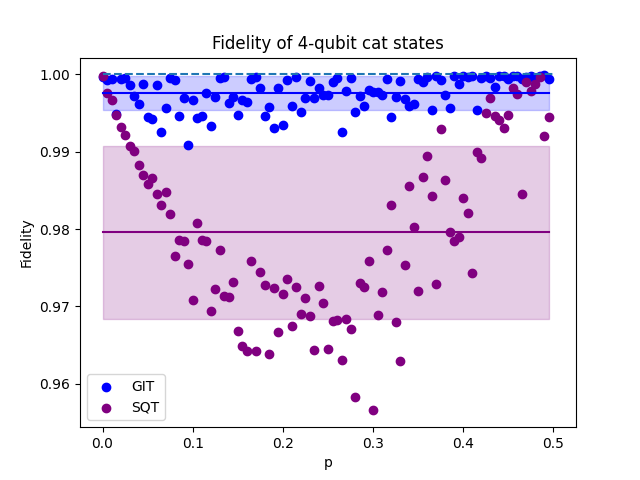}
\caption{Fidelity of 4-qubit cat states estimated via GIT (blue dots) and SQT (purple dots) using \cite{Kwiat-python}. Horizontal lines and shaded areas represent the mean fidelity value and standard deviation respectively for each tomographic scheme. The incomplete scheme performs better than its complete counterpart.}
\label{f:4q_Cat}
\end{figure}

 We studied cat-like states as a first example of this symmetry, which are given by $$|\psi_{p}\rangle=\sqrt{p}|0...0\rangle+\sqrt{1-p}|1...1\rangle$$for different values of the parameter $p$. In Fig. \ref{f:3q_Cat} and Fig. \ref{f:4q_Cat}, we used GIT to reconstruct these states considering $\lambda_\text{dc}=5\times10^{-5}$. The results obtained using a complete tomographic scheme are also included for comparison.


\begin{figure}[h]
\centering
\includegraphics[width=0.75\textwidth]{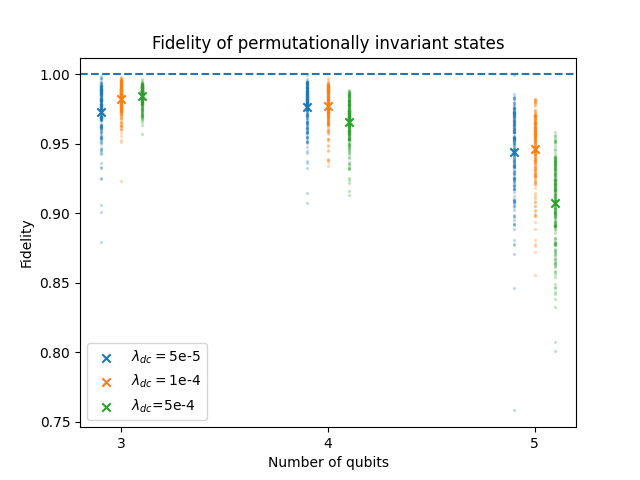}
\caption{Fidelities of 200 permutationally-invariant states estimated via GIT. We varied the number of qubits and used different values of dark counts $\lambda_{dc}$, increasing the noise in simulated frequency values as $\lambda_{dc}$ increases. Crosses indicate the mean fidelity value in each case.}
\label{f:Permutational}
\end{figure}

\begin{figure}[H]
\centering
\includegraphics[width=0.75\textwidth]{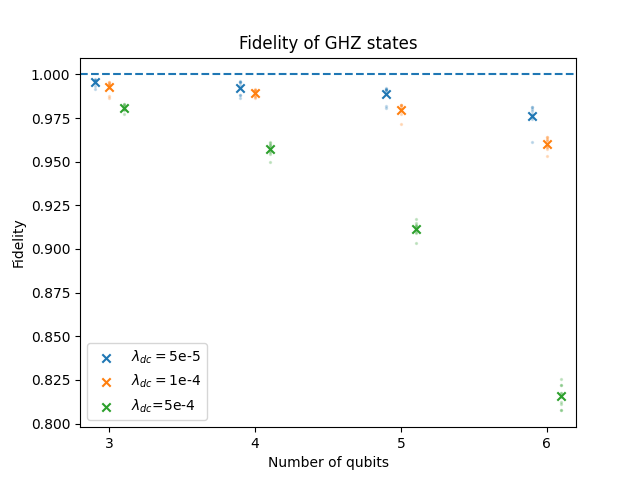}
\caption{Mean fidelity (crosses) of GHZ states estimated via GIT. We varied the number of qubits and used different values of dark counts $\lambda_{dc}$, increasing the noise in simulated frequencies values as $\lambda_{dc}$ increases. As expected, a greater number of $\lambda_{dc}$ leads to a worse fidelity.}
\label{f:GHZ}
\end{figure}

In Fig. \ref{f:Permutational} we show the mean fidelities of 200 permutationally-invariant reconstructed states. Another example of permutationally-invariant states are the Greenberger-Horne-Zeilinger (GHZ) states. In Fig. \ref{f:GHZ}, each cross indicates the mean fidelity of the estimated GHZ state using 10 different simulated sets of experimental frequencies.

\subsection{Werner states}
We now consider the family of {\textit{Werner states}} \cite{Werner1989,Eggeling2001}. In the case of a system of $N$ qubits, these states can be defined as those which are invariant under the action of the group
	\begin{equation}\label{werner_group}
	\mathcal{G}_{N}=\{U^{\otimes N} \,\,\mid\,\,U\in U(2)\}.
	\end{equation}
	That is, a Werner state satisfies the relation
	\begin{eqnarray}\label{sym_Werner}
	(U^{\otimes N})\rho_{\mathrm{(W)}}(U^{\otimes N})^{\dagger}=\rho_{\mathrm{(W)}},
	\end{eqnarray}
	for all unitary operators $U$ acting on the single-qubit space. Thus, these states have the symmetry defined by the action of the \textit{continuous group} $\mathcal{G}_{N}$. The generators of this group can be computed for all $N$. We illustrate the case of two qubits for simplicity (the general case follows similarly). Any unitary matrix $U$ can be written as $U= \exp(ia)$, where $a$ is an Hermitian $2\times 2$ matrix. Let $I$ be the $2\times 2$ identity matrix. Thus, $U\otimes U = \exp(ia)\otimes\exp(ia) = (\exp(ia)\otimes I)(I\otimes\exp(ia)) = \exp(ia\otimes I)\exp(I\otimes ia)=\exp(ia\otimes I + I \otimes ia)$. Now, we can write $a = \sum_{k=0}^{3}\alpha_{k}\sigma_{k}$, where the $\alpha$'s are real numbers, $\sigma_{0}=I$ and $\{\sigma_{k}\}_{k=1,2,3}$ are the Pauli matrices. Finally, $ia\otimes I + I \otimes ia = \sum_{k}\alpha_{k}(\sigma_{k}\otimes I + I \otimes \sigma_{k})$. This proves that the set $\{\sigma_{k}\otimes I + I \otimes \sigma_{k}\}_{k=0,1,2,3}$ is a basis for the generators of the group.

\begin{figure}[H]
\centering
\includegraphics[width=0.75\textwidth]{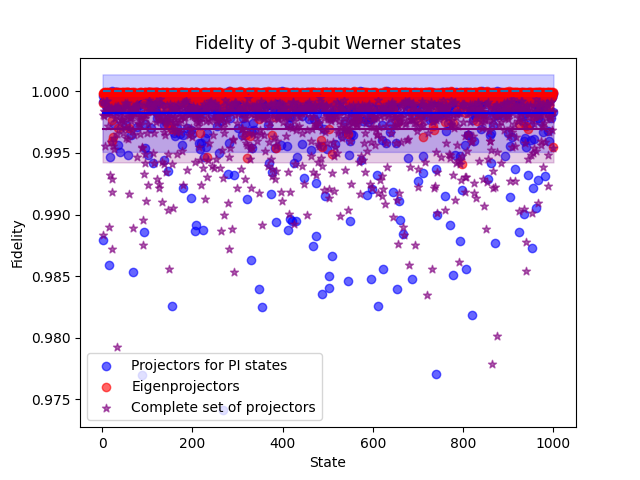}
\caption{Mean fidelity and standard deviation of 1000 random 3-qubit Werner states estimated via GIT. In this case we simulated measurements of different sets of projection operators: the set of 20 projectors used for the permutationally invariant (PI) states (blue dots) and 16 projectors (red dots) built from the eigenvectors of the orthogonal set of observables $S_i$ from Eq. \eqref{e:rho-parametrization}. Purple stars depict fidelities of estimations obtained via the SQT algorithm from \cite{Kwiat-python}.}
\label{f:Werner-PI}
\end{figure}

Once again, we focused on a potential quantum optics experiment. We considered two different measurement sets, giving the fidelities presented in Fig. \ref{f:Werner-PI}. As a first attempt, we simulated the projective measurements used for permutationally-invariant (PI) states. These projectors are more than the ones needed for quorum for this symmetry. However, this set is easy to implement in experimental setups and still implies fewer measurements compared to SQT. Moreover, the mean fidelity is greater in this case than in states estimated with SQT.
	
	We now examine a potentially harder experimental implementation. One way to reduce the number of measurements is measuring eigenprojectors of the orthogonal observables $\{S_i \}$ from Eq.\eqref{e:Parameterization} (i.e. operators projecting onto the eigenspace $E_i(\lambda_j)$ associated with the eigenvalue $\lambda_j$ of a given $S_i$). These projectors are fewer than the projectors used for PI states, but its amount is still greater than the number of parameters. As shown in Fig. \ref{f:Werner-PI}, the mean fidelity in this case is closer to 1 than in the approaches previously mentioned.

\begin{figure}[ht]
\centering
\includegraphics[width=0.75\textwidth]{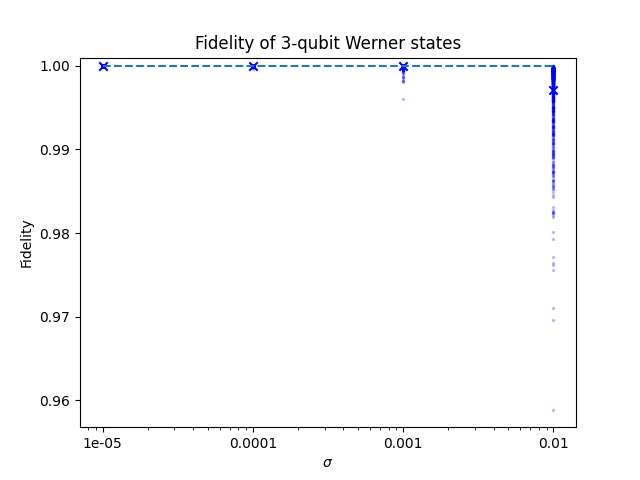}
\caption{Fidelity of random 3-qubit Werner states estimated via GIT. In this case we simulated measurements of the 5 orthogonal observables for the Werner symmetry implementing a Gaussian noise, with $\sigma$ ranging from $10^{-2}$ to $10^{-5}$. Each point represents a given state and the crosses indicate the mean fidelity for each value of $\sigma$. Even though the mean fidelity decreases as $\sigma$ grows, it is still over 0.996. For $\sigma=0.01$, states with fidelities greater than 0.95 still remain.}
\label{f:Werner}
\end{figure}

If possible, one could measure the set of orthogonal observables $\{S_i \}$ in Eq. \eqref{e:Parameterization} instead of projectors. In Fig. \ref{f:Werner}, we consider this situation with measurements affected by Gaussian noise. Simulated experimental measurements were generated adding a random number to each theoretical expectation value. These random numbers follow the normal distribution $\mathcal{N}(0, \sigma ^2)$, with $\sigma$ ranging from $10^{-2}$ to $10^{-5}$ in each case considered.

\subsection{Rotationally Invariant States}

In general, rotationally invariant states play a key role in many applications of quantum physics \cite{Vallone_RotationInvariant,DAmbrosio2012,Six_Photon_Rotationally_Invariant}. We will present some examples in which different types of rotational symmetry of the system under study are assumed to be known, and used as an advantage for state estimation.

\subsubsection{Global rotations}
In this case, we consider a symmetry whose generators are given by a single element set

\begin{equation}\label{e:Global_Rotations}
    R^{A}_{N} = \{
    \sigma_{z}\otimes I\otimes \cdots \otimes I + I \otimes \sigma_{z}\otimes \cdots \otimes I+\cdots +I\otimes I\otimes \cdots \otimes \sigma_{z}\}
\end{equation}

\noindent This is a continuous symmetry and represents arbitrary rotations of the system around the $z$ axis. The Bell state  $|\psi^{-}\rangle=\frac{1}{\sqrt{2}}(|01\rangle-|10\rangle)$ is an example of a rotationally-invariant state, given that it is invariant under the action of the bigger group $U\otimes U$, with $U$ an arbitrary unitary operator. Another example is a Werner state, in which the symmetry dictated by Eqn. \eqref{e:Global_Rotations} is represented by a subgroup of the group generated by \eqref{werner_group}.

Results are presented in Table \ref{t:RotA}. We considered a photonic setup and simulated the detection noise described in Section \ref{ejemplo:PI} with dc given by $\lambda_{dc}=5\times10^{-5}$. We computed the mean fidelities and mean state reconstruction times for states estimated using GIT. Reconstruction times and fidelities of states estimated via the SQT algorithm from \cite{Kwiat-python} are included for comparison.

\begin{table}
\centering
\begin{adjustbox}{max width=\textwidth}
\begin{tabular}{||c c c c c||}
 \hline
 No. of qubits & Fidelity for & Fidelity for & Reconstruction time & Reconstruction time\\
  & GIT & SQT & (GIT) & (SQT)\\ [0.5ex]
 \hline\hline
 3 & $0.99 \pm 0.01$ & $0.998 \pm 0.001$ & $< 1$ s& $\sim 30$ s\\ [0.5ex]
 \hline
 4 & $0.99 \pm 0.01$ & $0.997 \pm 0.002$ & $\sim 2.5$ s& $\sim 8$ min\\ [0.5ex]
 \hline
 5 & $0.99 \pm 0.01$ & $0.9908 \pm 0.0002$ & $\sim 22$ s & $\sim 2.5$ hours \\ [0.5ex]
 \hline
\end{tabular}
\end{adjustbox}
\caption{Mean fidelity of states invariant under global rotations reconstructed via GIT and SQT. We considered photonic devices with photon detection noise. Estimations of reconstruction times for each tomographic method are displayed. Times spent computing the orthogonal set of observables are not considered in this estimations.}
\label{t:RotA}
\end{table}

\subsubsection{Individual qubit rotations}

A more restricted family of states is given by arbitrary rotations of each qubit around the $z$ axis. This implies that each qubit can be rotated with a different angle. The group of individual qubit rotations is again continuous and its generators are given by
\begin{equation}
    R^{A}_{N} = \{
    \sigma_{z}\otimes I\otimes \cdots \otimes I, I \otimes \sigma_{z}\otimes \cdots \otimes I,\cdots,I\otimes I\otimes \cdots \otimes \sigma_{z}\}.
\end{equation}

Our results are shown in Table \ref{t:RotB}. Simulations were made under the same conditions of the global-rotation invariant states. We included the fidelities of states reconstructed via GIT and reconstruction times for GIT and SQT.

\begin{table}[H]
\centering
\begin{tabular}{||c c c c||}
 \hline
 No. of qubits & Fidelity for & Reconstruction time & Reconstruction time\\
  &  GIT & (GIT) & (SQT)\\ [0.5ex]
 \hline\hline
 3 & $0.98 \pm 0.02$ & $< 1$ s & $< 1$ s\\ [0.5ex]
 \hline
 4 &  $0.96 \pm 0.02$ & $< 1$ s & $\sim 15$ s\\ [0.5ex]
 \hline
 5 &  $0.95 \pm 0.03$ & $\sim 12$ s & $\sim 2.5$ hours\\ [0.5ex]
 \hline
\end{tabular}
\caption{Mean fidelity of states invariant under individual qubit rotations estimated via GIT. Measurements of 7, 15 and 31 projectors served as input for the estimation of states of 3, 4 and 5 qubits respectively. Estimations of reconstruction times for GIT and SQT are presented as well. Times spent computing the orthogonal set of observables are not considered in this estimations.}
\label{t:RotB}
\end{table}

\section{Conclusions}\label{s:Conclusions}

In this work, we have shown how to use \textit{a priori} knowledge of the symmetry of an otherwise unknown state to find a convenient parameterization for quantum state estimation purposes. This idea allows to reduce the number of independent parameters in a convex optimization formulation of the estimation problem, and leads to a concomitant reduction in the number of independent measurements as well.

Unlike previous methods, this approach works for arbitrary symmetries. This protocol can be potentially implemented on experimental scenarios and thus lower the difficulty level in practical implementations. Moreover, symmetries are taken into account in a relatively simple algorithmic way that does not depend on complex mathematical properties of the groups involved. We think that the results presented in this work are a step forward in the development of state estimation protocols, which can be used to reduce both the computational and experimental complexity of the existing methods.

\providecommand{\noopsort}[1]{}\providecommand{\singleletter}[1]{#1}%

\end{document}